%% file: wz_flow_0717.tex
\documentclass[preprint,12pt]{elsarticle}

\usepackage{amsmath}
\usepackage{amssymb}
\usepackage{amsfonts}
\usepackage{graphics}
\usepackage{graphicx}
\usepackage{amscd}
\usepackage{amsfonts}
\usepackage{epsf}
\usepackage{epsfig}
\usepackage{color}

 \usepackage[utf8]{inputenc} 

\biboptions{numbers,sort&compress}

\usepackage{etoolbox}
\makeatletter
\patchcmd{\ps@pprintTitle}
  {\let\@oddhead\@empty}
  {\def\@oddhead{\mbox{}\hfill \footnotesize{MISC-2019-01, KUNS-2756, UTCCS-P-116}}}
  {}{}
\makeatother

\newcommand{\n}{\nonumber}

\newcommand{\eref}[1]{(\ref{#1})}

\newcommand{\aeq}{& \hspace{-2mm} =\hspace{-2mm} &}
\allowdisplaybreaks[1]

\begin{document}

\begin{frontmatter}

\title{Supersymmetric gradient flow in the Wess-Zumino model} 

\author{Daisuke Kadoh}
\address{Department of Physics, Faculty of Science, Chulalongkorn University,
Bangkok 10330, Thailand\\  Research and Educational Center for Natural Sciences, Keio University, Yokohama 223-8521, Japan}
\ead{kadoh@keio.jp}

\author{Kengo Kikuchi}
\address{Maskawa Institute for Science and Culture, Kyoto Sangyo University, Kyoto 603-8555, Japan\\
Department of Physics, Kyoto University, Kyoto 606-8502, Japan}
\ead{kengo@yukawa.kyoto-u.ac.jp}

\author{Naoya Ukita}
\address{Center for Computational Sciences, University of Tsukuba, Tsukuba, Ibaraki 305-8577, Japan}
\ead{ukita@ccs.tsukuba.ac.jp}

\begin{abstract}
We propose a supersymmetric gradient flow equation in the four-dimensional Wess-Zumino model. 
The flow is constructed in two ways.
One is based on the off-shell component fields and the other is based on the superfield formalism, 
in which the same result is provided.
The obtained flow is supersymmetric because the flow time derivative and the supersymmetry 
transformation commute with each other. Solving the equation, we find that 
it has a damping oscillation with the flow time for nonzero mass,  which is different from the Yang-Mills flow.  
The on-shell flow equation is also discussed.
\end{abstract}

\begin{keyword}
lattice, supersymmetry
\end{keyword}

\end{frontmatter}

\input{sec1}

\input{sec2}

\input{sec3}

\input{sec4}

\input{sec5}

\section*{Acknowledgement}

This work was supported by JSPS KAKENHI Grants No.~JP16K05328, No.~JP16K13798, No.~JP18K13546, No.~JP19K03853 the CUniverse research promotion project of Chulalongkorn University (Grant No.~CUAASC). 
We thank Iwanami Fujukai for the financial aid.

\appendix

\input{app}


\providecommand{\href}[2]{#2}\begingroup\raggedright\endgroup

\end{document}

%% file: sec1.tex
\section{Introduction} 
\label{sec:introduction}

Gradient flow \cite{Narayanan:2006rf, Luscher:2010iy} has been widely accepted 
as a new method in lattice field theory and related research areas including supersymmetry (SUSY). 
In a Yang-Mills flow, any correlation function is ultraviolet (UV) finite at nonzero flow time 
once four-dimensional Yang-Mills theory is renormalized \cite{Luscher:2011bx}. 
The UV finiteness holds even for a QCD flow 
with an additional renormalization of the time-dependent quarks \cite{Luscher:2013cpa, Hieda:2016xpq}.   
This property of the flow leads to 
a lot of interesting applications 
such as a proper definition of lattice energy momentum 
tensor \cite{Suzuki:2013gza, DelDebbio:2013zaa, Asakawa:2013laa, Makino:2014taa, 
Taniguchi:2016ofw, Kitazawa:2017qab, Yanagihara:2018qqg, Harlander:2018zpi, Iritani:2018idk}. 
The gradient flow approach is also useful in studying the nonlinear sigma 
model \cite{Makino:2014sta, Aoki:2014dxa, Makino:2014cxa, Bietenholz:2018agd}, 
nonperturbative renormalization group 
\cite{Yamamura:2015kva,  Makino:2018rys,  Abe:2018zdc, Carosso:2018bmz, Sonoda:2019ibh}, 
and a theory with anti-de Sitter geometries \cite{Aoki:2015dla, Aoki:2016ohw, Aoki:2016env, Aoki:2017bru, Aoki:2017uce}. 
Other interesting applications are in the references 
\cite{Endo:2015iea, Fujikawa:2016qis, Hieda:2016lly, Taniguchi:2016tjc, Morikawa:2018fek, Suzuki:2018vfs}.

There have been various attempts to apply the gradient flow to SUSY theories so far. 
In super Yang-Mills (SYM), the most naive approach is to use a non-SUSY flow, 
which consists of the Yang-Mills flow and an adjoint matter flow \cite{Luscher:2013cpa} 
although SUSY is broken at a nonzero flow time.  
From this point of view,  a lattice simulation of ${\cal N}=1$ SYM 
has been carried out in \cite{Bergner:2019dim} and the regularization 
independent definition of the supercurrent in ${\cal N}=1,2$ SYM has been given in \cite{Hieda:2017sqq, Kasai:2018koz}.

A different approach can be taken, which uses a flow keeping SUSY at a nonzero flow time. 
Such a SUSY flow
has been proposed in the superfield formalism of  ${\cal N}=1$ SYM \cite{Kikuchi:2014rla}. \footnote{
In the context of the Langevin equation, 
a flow equation for ${\cal N}=1$ SYM was discussed in \cite{Nakazawa:2003zf,Nakazawa:2003tz}.
}
The SUSY flow equation is also given for the component fields of the Wess-Zumino gauge 
in a gauge covariant manner~\cite{Kadoh:2018qwg}. The obtained flow is supersymmetric in a sense that the flow time derivative and the super transformations commute up to a gauge transformation.
The flow equation of supersymmetric O(N) nonlinear sigma model in two dimensions is also studied in \cite{Aoki:2017iwi}.

The Wess-Zumino model provides a good testing ground to study 
the  renormalization property of the SUSY theories.
The gauge symmetry plays a crucial role to prove the UV finiteness in the Yang-Mills flow. 
As natural questions, 
one might ask how SUSY works in the SUSY flows and what kind of influence 
the nonrenormalization theorem has for the flow theory.
Constructing a SUSY flow for the Wess-Zumino model, 
a  deep understanding of the mechanism that leads to the UV finiteness 
of the SUSY flows could be obtained.

In this paper, we derive a SUSY flow of  the four-dimensional Wess-Zumino model, 
which is referred as Wess-Zumino flow in this paper, 
and derive its formal solution.
We give two ways of constructing the Wess-Zumino flow.
One way is to use the component fields of the model directly, 
and the other way is to use the superfield formalism. 
They give the same result.
Solving the Wess-Zumino flow, 
we find that the solutions behave as damping oscillations 
with respect to the flow time for nonzero mass,
which is different from the Yang-Mills flow.

This paper is organized as follows.  In Sec.\ref{sec:wess-zumino}, 
we give the brief review of Wess-Zumino model in four dimensions. 
In Sec.\ref{sec:wz_flow}, we present two methods of constructing 
the Wess-Zumino flow equation.
We first present the results in Sec.\ref{sec:5d_susy}. 
The Wess-Zumino flow is constructed 
with the component fields in Sec.\ref{sec:derivation_component_field}  and 
with the superfield formalism in Sec.\ref{sec:derivation_superfield_formalism}.
The on-shell flow is also discussed in Sec.\ref{sec:onshell_flow}.
The formal solutions of the Wess-Zumino flow are given in Sec.\ref{sec:solution}. 
We summarize our results in Sec.\ref{sec:summary}. 
The convention used in this paper is shown in \ref{sec:notation}.

%% file: sec2.tex
\section{Wess-Zumino model} 
\label{sec:wess-zumino}
We make a brief review of the Wess-Zumino model, which 
is the simplest supersymmetric theory 
made of a complex scalar $A(x)$, 
Weyl spinors $\psi_\alpha (x)$, $\bar\psi_{\dot\alpha} (x)$, and a complex auxiliary field $F(x)$.

The action in Euclidean space is given by
\begin{eqnarray}
&& S =\int d^4 x \bigg\{ |\partial_\mu A |^2 
+ i\psi \sigma_{\mu} \partial_\mu \bar{\psi}
+ |F|^2 -i \big( F (m A+gA^2) + h.c. \big)  \nonumber\\
&& \hspace{2cm} 
 +\frac{1}{2} \psi \psi (m+2gA) +\frac{1}{2} \bar\psi \bar\psi (m +2 g A)^*
\bigg\},
\label{action_component}
\end{eqnarray} 
where a real and non-negative mass $m \ge 0$ and $g \in \mathbb{C}$, which can be chosen by a phase rotation of the fields  without loss of generality.
The off-shell SUSY transformation that makes the action (\ref{action_component}) invariant is defined as 
\begin{eqnarray}
\begin{split}
&\delta_\xi A(x)=  \xi \psi (x), \\
&\delta_\xi A^*(x)= \bar\xi  \bar{\psi} (x), \\
&\delta_\xi \psi (x)=i  \sigma_{\mu}\bar\xi \partial_\mu A(x)+i \xi F(x), \\
&\delta_\xi \bar{\psi}(x)=i  \bar{\sigma}_{\mu} \xi \partial_\mu A^{*}(x)+i \bar\xi F^*(x), \\
&\delta_\xi F(x) = \bar\xi  \bar{\sigma}_\mu \partial_\mu\psi (x), \\
&\delta_\xi F^* (x)= \xi \sigma_{\mu} \partial_{\mu}\bar{\psi} (x),
\end{split}
\label{off-shell_susy}
\end{eqnarray}
where $\xi_\alpha$ and $\bar\xi_{\dot\beta}$ are two anticommuting parameters.
The off-shell transformation satisfies
\begin{eqnarray}
[\delta_\xi, \delta_\eta ] = -i\left(\bar\xi\bar\sigma_{\mu}\eta+\xi\sigma_{\mu}\bar\eta\right)\partial_{\mu},
\label{susy-agebra}
\end{eqnarray}
which is a well-known relation derived from the SUSY algebra.

The on-shell action is obtained as
\begin{eqnarray}
&& S_{\rm on-shell} = \int d^4 x \bigg\{ |\partial_\mu A |^2 +| mA + g A^2|^2 
+ i\psi \sigma_{\mu} \partial_\mu \bar{\psi}\nonumber\\
&& \hspace{3cm} 
 + \frac{1}{2} \psi \psi (m + 2 g A) +\frac{1}{2} \bar\psi \bar\psi (m +2 g A)^* \bigg\},
 \label{on-shell_action}
\end{eqnarray}
integrating the auxiliary field $F$ of the off-shell one (\ref{action_component}).
The action  (\ref{on-shell_action})  is invariant under the on-shell SUSY transformation,
\begin{eqnarray}
\begin{split}
&\delta^\prime_\xi A(x)=  \xi \psi (x), \\
&\delta^\prime_\xi A^*(x)= \bar\xi  \bar{\psi} (x), \\
&\delta^\prime_\xi \psi (x)=i  \sigma_{\mu}\bar\xi \partial_\mu A(x) -  \xi (mA^*+g^*A^{*2})(x), \\
&\delta^\prime_\xi \bar{\psi}(x)=i  \bar{\sigma}_{\mu} \xi \partial_\mu A^{*}(x) - \bar\xi  (m A +g A^{2})(x).
\end{split}
\label{on-shell_susy}
\end{eqnarray}
Note that (\ref{on-shell_susy}) are the first four transformations of (\ref{off-shell_susy})
replacing $F \rightarrow  i (mA^*+g^*A^{*2})$ and $F^* \rightarrow i (m A+gA^{2})$. 

The off-shell SUSY theory is also easily defined using the superfield formalism. 
Suppose that $\theta_\alpha$ and $\bar \theta_{\dot\alpha}$ are two global Grassmann parameters.
Superfield is then defined by a function ${\cal F}(x, \theta, \bar\theta) $ 
whose SUSY transformation is given by
\begin{eqnarray}
\delta_\xi {\cal F} (x,\theta,\bar\theta) = \frac{1}{\sqrt{2}} (\xi Q +\bar \xi \bar Q) {\cal F} (x,\theta,\bar\theta)
\label{susy_superfield}
\end{eqnarray}
where $Q_\alpha$ and $\bar Q_{\dot\alpha}$ are differential operators,
\begin{eqnarray}
&&Q_\alpha=\frac{\partial}{\partial \theta^{\alpha}}
-i(\sigma_{\mu})_{\alpha\dot\alpha}\bar\theta^{\dot\alpha}\partial_{\mu},\\
&&\bar{Q}_{\dot\alpha}=-\frac{\partial}{\partial \bar\theta^{\dot\alpha}} 
+i\theta^{\alpha}(\sigma_{\mu})_{\alpha\dot\alpha}\partial_{\mu}.
\end{eqnarray}
For later use, we introduce other differential operators,
\begin{eqnarray}
&&D_\alpha=\frac{\partial}{\partial\theta^\alpha} 
+i(\sigma_{\mu})_{\alpha\dot\alpha}\bar{\theta}^{\dot\alpha}\partial_{\mu},\\
&&\bar D_{\dot\alpha}=-\frac{\partial}{\partial \bar{\theta}^{\dot\alpha}}
-i\theta^{\alpha}(\sigma_{\mu})_{\alpha\dot\alpha}\partial_{\mu},
\label{DD}
\end{eqnarray}
which are covariant under SUSY transformation (\ref{susy_superfield}) because 
\begin{eqnarray}
&&\{ Q_\alpha, \bar Q_{\dot\alpha} \} = - \{ D_\alpha, \bar D_{\dot\alpha} \} 
= 2i(\sigma_{\mu})_{\alpha\dot\alpha}\partial_{\mu},
\end{eqnarray}
and the other commutation relations are 0.

The Wess-Zumino model is given
by chiral and antichiral superfields $\Phi(x,\theta,\bar\theta)$  and $\bar\Phi(x,\theta,\bar\theta)$,
which satisfy
\begin{eqnarray}
\bar D_{\dot\alpha} \Phi=D_{\alpha} \bar\Phi=0.
\label{super_chiral}
\end{eqnarray}
The  $\theta$ and $\bar\theta$ expansion of the chiral superfields can easily be written in terms of 
$y_\mu=x_\mu + i\theta\sigma_\mu \bar{\theta}$ and $\bar y_\mu=x_\mu - i\theta\sigma_\mu \bar{\theta}$ 
because,  for instance, $\bar D_{\dot\alpha}=-\frac{\partial}{\partial \bar{\theta}^{\dot\alpha}}$ in the $y$ coordinate.
We thus have 
\begin{eqnarray}
\begin{split}
&\Phi(y,\theta) =A(y)+\sqrt{2}\theta\psi(y)+i\theta\theta F(y),\\
& \bar \Phi( \bar y,\bar{\theta}) = A^*(\bar y)+\sqrt{2}\bar\theta\bar\psi(\bar y)+i\bar\theta\bar\theta F^*(\bar y). 
\end{split}
\label{Phi_expansion}
\end{eqnarray}
The off-shell SUSY transformation for the component fields (\ref{off-shell_susy}) are reproduced 
from  the definition (\ref{susy_superfield}) with the expansion (\ref{Phi_expansion}).

The off-shell action  (\ref{action_component}) can also be expressed as
\begin{eqnarray}
&& S=-\int d^4 x \bigg\{ {\bar \Phi  \Phi} |_{\theta\theta \bar\theta\bar\theta} 
+ W(\Phi)|_{\theta\theta} +  W^*(\bar\Phi)|_{\bar\theta\bar\theta} \bigg\},
\label{supeprfield_action}
\end{eqnarray}
where 
\begin{eqnarray}
\begin{split}
&W(\Phi)=\frac{1}{2}m\Phi^2+\frac{1}{3}g\Phi^3.\label{w}
\end{split}
\end{eqnarray}
From the construction presented above, it is obvious that the superfield action (\ref{supeprfield_action})
is invariant 
under the off-shell SUSY transformation (\ref{susy_superfield}).

%% file: sec3.tex
\section{Wess-Zumino flow} 
\label{sec:wz_flow}

We construct a supersymmetric flow equation in the Wess-Zumino model.  
The flow equation is derived in two ways; one is based on the off-shell component fields 
as shown in Sec.\ref{sec:derivation_component_field}
and the other is based on the superfield formalism as seen 
in Sec.\ref{sec:derivation_superfield_formalism}.
We find that they give the same result.

\subsection{4+1-dimensional supersymmetry and supersymmetric flow}
\label{sec:5d_susy}
We introduce a flow time  $t\ (\ge 0)$ and consider the time-dependent bosonic fields 
$\phi(t,x)$, $\bar\phi(t,x)$, $G(t,x)$, $\bar G(t,x)$ $\in \mathbb{C}$ 
and spinors $\chi(t,x), \bar\chi(t,x)$.
The component fields of the Wess-Zumino model are replaced by those fields as follows:
\begin{eqnarray}
\begin{split}
  A(x) \ \ &\rightarrow \ \ \phi(t,x) \\
  A^*(x)  \ \  &\rightarrow  \ \  \bar\phi(t,x) \\
  \psi(x)   \ \  &\rightarrow  \ \  \chi(t,x) \\
  \bar\psi(x)  \ \  &\rightarrow  \ \  \bar \chi(t,x) \\
  F(x)  \ \  &\rightarrow   \ \ G(t,x) \\
  F^*(x)  \ \  &\rightarrow \ \  \bar G(t,x), \\
\end{split}
\label{flow_fields}
\end{eqnarray}
with boundary conditions, 
\begin{eqnarray}
\begin{split}
 & (\phi(t,x), \chi(t,x),G(t,x))|_{t=0} = (A(x),\psi(x),F(x)), \\ 
  & (\bar \phi(t,x), \bar\chi(t,x),\bar G(t,x))|_{t=0} = (A^*(x),\bar\psi(x),F^*(x)). \\ 
\end{split}
\label{initial}
\end{eqnarray}
Note that $\bar\phi$ and $\bar G$ are no longer the complex conjugates of $\phi$ and $G$, respectively, 
for nonzero flow time.

For the flowed fields, 
4+1-dimensional SUSY transformation can be defined 
by replacing the fields of \eref{off-shell_susy} according to (\ref{flow_fields}),
\begin{eqnarray}
\begin{split}
&\delta_\xi \phi =  \xi \chi, \\
&\delta_\xi \bar\phi = \bar\xi  \bar{\chi},  \\
&\delta_\xi \chi =i  \sigma_{\mu}\bar\xi \partial_\mu \phi +i \xi G,  \\
&\delta_\xi \bar{\chi} =i  \bar{\sigma}_{\mu} \xi \partial_\mu \bar \phi +i \bar\xi \bar G, \\
&\delta_\xi G  = \bar\xi  \bar{\sigma}_\mu \partial_\mu\chi,  \\
&\delta_\xi \bar G = \xi \sigma_{\mu} \partial_{\mu}\bar{\chi}, 
\end{split}
\label{5dsusy}
\end{eqnarray} 
where $\xi$ and $\bar\xi$ are $t$-independent parameters.  

It is shown that
a supersymmetric flow equation is given by
\begin{eqnarray}
&& \partial_t \phi  = \Box \phi +im\bar G +g^*\left( 2 i \bar\phi \bar G -\bar\chi \bar\chi \right), 
\label{flow_phi}
\\
&&\partial_t \bar \phi  = \Box \bar\phi  +imG  +g \left( 2 i \phi G -\chi  \chi \right), 
\label{flow_barphi}
\\
&& \partial_t \chi   = \Box\chi  +i\sigma_\mu\partial_\mu \left(m \bar\chi+2g^* \bar\phi \bar\chi \right),  
\label{flow_chi}
\\
&& \partial_t \bar\chi  = \Box\bar\chi +i\bar\sigma_\mu\partial_\mu \left(m \chi + 2 g \phi \chi \right),   
\label{flow_barchi}
\\
&& \partial_t G = \Box  G-i \Box\left( m \bar \phi +g^* \bar \phi^{2} \right),
\label{flow_G}
\\
&& \partial_t \bar G = \Box \bar G -i \Box \left(m \phi  +g \phi^{2}  \right),
\label{flow_barG}
\end{eqnarray}
where $\Box=\sum_\mu \partial_\mu \partial_\mu$.

The Wess-Zumino flow tells us that each component field does not flow independently  
but mixes with other fields to keep SUSY. 
The flowed fields $G$ and $\bar G$ are no longer auxiliary fields because 
derivative terms are in (\ref{flow_G}) and (\ref{flow_barG}).
It is possible to show that
\begin{eqnarray}
\left[\partial_t,\delta_\xi \right] = 0,
\label{consistency_condition}
\end{eqnarray}
which means that SUSY is kept at nonzero flow time. 
As we see in the next two sections, 
(\ref{consistency_condition}) can also be easily confirmed from the construction of 
the Wess-Zumino flow equation.

\subsection{Derivation of the Wess-Zumino flow in component fields}
\label{sec:derivation_component_field}

We begin with considering a derivative of $S$ with respect to $A(x)$. 
Since $\delta S / \delta A(x)$ has $\Box A^*(x)$, 
a gradient flow for $\phi(t,x)$ as a diffusion equation 
$\partial_t \phi \simeq \Box \phi $
should be defined by
\begin{eqnarray}
\partial_t \phi(t,x) = -\left. \frac{\delta S}{\delta A^*(x)} \right|
_{{\rm fields} \ \rightarrow \ {\rm flowed \ fields}},
\label{flow_phi_def}
\end{eqnarray}
where $X |_{{\rm fields} \ \rightarrow \ {\rm flowed \ fields}}$ 
means that the field variables in $X$ are replaced according to (\ref{flow_fields}).
The first flow equation (\ref{flow_phi}) is obtained from (\ref{flow_phi_def}).
Similarly, (\ref{flow_barphi}) is derived 
from a gradient flow equation as  (\ref{flow_phi_def}) by replacing  $\partial_t \phi$ and $\delta A^*$ by $\partial_t \bar\phi$ and $\delta A$, respectively.

Suppose that (\ref{consistency_condition}) holds 
for $\phi$. 
Then the L.H.S. of  (\ref{flow_phi}) becomes
\begin{eqnarray}
\delta_\xi \partial_t \phi(t,x) =\partial_t  \delta_\xi \phi(t,x) = \xi \partial_t \chi(t,x),
\label{flow_chi_lhs}
\end{eqnarray}
while the SUSY transformation of the R.H.S. of  (\ref{flow_phi}) is 
\begin{eqnarray}
\delta_\xi ({\rm R.H.S. \ of} \,  (\ref{flow_phi})) = 
\xi \left( \Box\chi  +i\sigma_\mu\partial_\mu \left(m\bar\chi+2g^* \bar\phi \bar\chi \right) \right).
\label{flow_chi_rhs}
\end{eqnarray}
Since (\ref{flow_chi_lhs}) coincides with  (\ref{flow_chi_rhs}),  
we obtain  (\ref{flow_chi}). We can also find  (\ref{flow_barchi}) 
assuming  (\ref{consistency_condition}) for $\bar\phi$ as well.

The flow equations for $G$ and $\bar G$ are derived in the same manner.
If  (\ref{consistency_condition}) holds for $\chi$ and $\bar \chi$, 
we immediately find (\ref{flow_G}) and (\ref{flow_barG}) by performing 
 the SUSY transformation of the flow equations for $\chi$ and $\bar \chi$.

Once the flow equations are given  for the scalar fields, 
we found that those for the other fields can be constructed
by repeating the SUSY transformation (\ref{5dsusy}).
Since we then assumed  (\ref{consistency_condition}) 
for $\phi,\bar\phi, \chi$ and $\bar \chi$, it is obvious that
 the obtained flows satisfy  (\ref{consistency_condition}) for them.
So all we have to do is check whether   (\ref{consistency_condition}) holds for $G$ and $\bar G$ or not.

It is enough to consider two cases:  (a) $\delta_\xi$ for $\bar \xi=0$ (and $\xi\neq 0$) and (b) $\delta_\xi$ for  $\xi=0$ (and $\bar \xi\neq0$) since a general $\delta_\xi$ is the summation of (a) and (b).
We now have $\delta_\xi \bar\phi =\delta_\xi G=0$ for $ \bar\xi=0$.
So it can be immediately shown that $[\partial_t, \delta_\xi] G=0$ for $ \bar\xi=0$. 
Moreover, one can show that  
\begin{eqnarray}
[\partial_t,\delta_\xi] \eta G - [\partial_t,\delta_\eta]  \xi G = 0,
\label{consistency_on_G}
\end{eqnarray}
from  the SUSY transformation of $\chi$
using $[\partial_t,\delta_\xi]= 0$ for $\phi, \chi$ and $[\partial_t, [\delta_\xi,\delta_\eta]]=0$. 
From (\ref{consistency_on_G}), we also confirm that 
$[\partial_t, \delta_\xi] G=0$ for $\xi=0$. 
We thus obtain $[\partial_t,\delta_\xi] G=0$ for a general $\delta_\xi$.  
Repeating the same argument, (\ref{consistency_condition}) is also true for $\bar G$.

\subsection{Derivation of Wess-Zumino flow in superfield formalism}
\label{sec:derivation_superfield_formalism}

The flowed superfields are given by replacing 
\begin{eqnarray}
\Phi(z) \rightarrow \Psi(t,z), \quad  \bar\Phi(z) \rightarrow \bar\Psi(t,z), 
\label{flow_superfields}
\end{eqnarray}
with $z=(x,\theta,\bar\theta)$.  
Suppose that 
\begin{eqnarray}
\Psi(t,z)|_{t=0}=\Phi(z), \quad  \bar\Psi(t,z)|_{t=0} =  \bar\Phi(z)
\end{eqnarray}
as an initial condition
and the SUSY transformation of $\Psi(t,z)$ and $\bar\Psi(t,z)$ is defined by (\ref{susy_superfield}).

The gradient flow should be given 
such that $\Phi(t,z)$ and $\bar\Phi(t,z)$ are chiral and antichiral superfields 
satisfying (\ref{super_chiral}). 
The field variation of the chiral superfield is defined as
\begin{eqnarray}
\frac{\delta}{\delta \bar \Phi(\bar y,\theta)} \bar \Phi(\bar y^\prime,\theta^\prime) 
= \delta^4 (\bar y- \bar y^\prime) \delta^2(\theta-\theta^\prime).
\end{eqnarray}
It can be shown that 
\begin{eqnarray}
\frac{\delta S}{\delta \bar \Phi(x, \theta, \bar \theta)}
\aeq \frac{1}{4}DD\Phi(x, \theta, \bar \theta)
 - \frac{\partial W^{*}(\bar\Phi(x, \theta, \bar \theta))}{\partial \bar\Phi(x, \theta, \bar \theta)}.
\label{delta_S}
\end{eqnarray}
Although it is natural to use $\delta S/\delta \bar\Phi$ for a gradient flow for $\Phi$, 
such a derivative does not satisfy the supersymmetric chiral condition (\ref{super_chiral}) for $\Phi$. 

It is possible to keep the condition (\ref{super_chiral}) multiplying $\delta S/{\delta \bar \Phi}$ by $\bar D \bar D$. 
Thus a proper flow equation is  
\begin{eqnarray}
\partial_t \Psi(t, z)  =
\frac{1}{4}\bar D \bar D \left.\frac{\delta S}{\delta \bar\Phi(z)}
\right|_{\Phi(z),\bar\Phi(z)  \ \rightarrow \Psi(t,z),\bar\Psi(t,z)},
\label{flow_eq_Psi}
\end{eqnarray}
and similarly  
\begin{eqnarray}
\partial_t \bar\Psi(t, z)  =
\frac{1}{4} D D \left.\frac{\delta S}{\delta \Phi(z)}
\right|_{\Phi(z),\bar\Phi(z)  \ \rightarrow \Psi(t,z),\bar\Psi(t,z)}.
\label{flow_eq_barPsi}
\end{eqnarray}
Since $\bar D \bar D DD=16 \Box$, we have
\begin{eqnarray}
\begin{split}
& \partial_t \Psi =\Box\Psi - \frac{1}{4}\bar{D}\bar{D} W^{\prime *}(\bar{\Psi}),\\
& \partial_t \bar \Psi  = \Box\bar\Psi - \frac{1}{4} D DW'(\Psi),
\end{split}
\label{super_flow_superfield}
\end{eqnarray}
where $W^\prime(x)= \partial W(x)/\partial x$. The supersymmetric chiral condition (\ref{super_chiral})  is actually kept for any nonzero flow time
because, noticing $D^3=\bar D^3=0$ and $[D,\partial_t]=[\bar D,\partial_t]=0$,
\begin{eqnarray}
\partial_t( \bar D_{\dot \alpha} \Psi(t,x) ) = \partial_t( D_{\alpha} \bar \Psi(t,x) )=0,
\end{eqnarray}
with $\bar D_{\dot \alpha} \Psi(t=0,x) = D_{\alpha} \bar \Psi(t=0,x)=0$.

The definitions of the gradient flow (\ref{flow_eq_Psi}) and (\ref{flow_eq_barPsi})
are consistent with the SUSY transformation given by (\ref{susy_superfield}) because $[Q,\partial_t]=[\bar Q,\partial_t]=0$. So the commutation relation (\ref{consistency_condition}) is manifestly satisfied.

Since the flowed superfields obey the supersymmetric chiral condition  (\ref{super_chiral}), 
they can also be expanded as 
\begin{eqnarray}
\begin{split}
& \Psi(t,y,\theta) =\phi(t,y)+\sqrt{2}\theta\chi(t,y)+i\theta\theta G(t,y),\\
& \bar \Psi(t,\bar y,\bar{\theta}) = \bar\phi(t,\bar y)+\sqrt{2}\bar\theta\bar\chi(t,\bar y)+i\bar\theta\bar\theta \bar G(t, \bar y).
\end{split}
\label{component_expression}
\end{eqnarray}
Substituting these expansions into (\ref{super_flow_superfield}), we find that 
the same flow equations as (\ref{flow_phi})--(\ref{flow_barG}) are obtained.

\subsection{The on-shell flow}\label{sec:onshell_flow}

The relation (\ref{consistency_condition}) is shown to be satisfied for the off-shell supersymmetric gradient flow.  
 We mention an on-shell case in which the auxiliary field is integrated out. 

We consider an on-shell flow by replacing $G$ and $\bar G$ of 
(\ref{flow_phi})--(\ref{flow_barchi})
as
\begin{eqnarray}
G =  i (m \bar\phi + g^* \bar\phi^{2}), \qquad \bar G = i (m \phi + g \phi^{2}),
\label{on-shell_condition}
\end{eqnarray}
which are the equations of motion of $F$ and $F^*$ at $t=0$. Here we do not consider the flow equation of $G$ and $\bar G$.
An on-shell SUSY transformation $\delta^\prime_\xi$ for the flowed fields is given by \eref{on-shell_susy} with the replacement \eref{flow_fields}.

The commutation relation between the flow derivative and
 the on-shell SUSY transformation does not vanish in general 
but is proportional to $\delta S/\delta h$ for $h=\psi, A, \bar\psi, A^*$. 
For instance, 
\begin{eqnarray}
[ \partial_t, \delta^{\prime}_{\xi}] \phi 
= W^{''*}(\bar\phi) \xi \left. \frac{\delta S}{\delta \psi}\right|_{{\rm fields} \rightarrow {\rm {flowed~fields }}}.
\end{eqnarray}
One can easily show that the commutators
for other fields do not also vanish but satisfy the similar relations.

%% file: sec4.tex
\section{Formal solution of Wess-Zumino flow} 
\label{sec:solution}

The flowed chiral and antichiral superfields are directly coupled even in the linear part of the Wess-Zumino flow, 
\begin{eqnarray}
\partial_t 
\left( \hspace{-1.5 mm}
\begin{array}{c}
 \Psi_0 \vspace{3mm}  \\
 \bar \Psi_0  
\end{array}
\hspace{-1.5 mm} \right)
=
\left( \hspace{-1.5 mm}
\begin{array}{cc}
 \Box &   -\frac{m}{4}\bar{D}\bar{D} \vspace{3mm} \\
 -\frac{m}{4} D D  &  \Box
\end{array}
\hspace{-1.5 mm} \right)
\left( \hspace{-1.5 mm}
\begin{array}{c}
 \Psi_0 \vspace{3mm}  \\
 \bar \Psi_0  
\end{array}
\hspace{-1.5 mm} \right),
\label{wz_flow_superfield_linear_part}
\end{eqnarray}
where the suffix 0 means they are solutions to the linear part of the flow equation.

To solve the formal solution of the Wess-Zumino flow,  
let us move on to a basis that 
diagonalizes the matrix of (\ref{wz_flow_superfield_linear_part}) as
\begin{eqnarray}
\left( \hspace{-1.5 mm}
\begin{array}{c}
 \Pi_{+} \vspace{3mm}  \\
 \Pi_{-}  
\end{array}
\hspace{-1.5 mm} \right)
=
\frac{1}{\sqrt{2}}
\left( \hspace{-1.5 mm}
\begin{array}{cc}
 \frac{i}{4}\frac{DD}{\sqrt{-\Box}}   & 1  \vspace{3mm} \\
 -\frac{i}{4}\frac{DD}{\sqrt{-\Box}}   & 1
\end{array}
\hspace{-1.5 mm} \right)
\left( \hspace{-1.5 mm}
\begin{array}{c}
 \Psi \vspace{3mm}  \\
 \bar \Psi  
\end{array}
\hspace{-1.5 mm} \right).
\label{pi_superfield}
\end{eqnarray}
Then the Wess-Zumino flow equation is given in terms of $\Pi_{+}$ and  $\Pi_{-}$,
\begin{eqnarray}
\begin{split}
& \partial_t \Pi_{\pm} =\left( \Box \pm im\sqrt{-\Box}\right )\Pi_{\pm} +R_{\pm},
\end{split}
\label{super_flow_pi_field}
\end{eqnarray}
where
\begin{eqnarray}
R_{\pm} = \pm\frac{ig^*\sqrt{-\Box}}{\sqrt{2}}\bar\Psi^2
         -\frac{g}{4\sqrt{2}}DD \Psi^2.
\end{eqnarray}
Note that the initial conditions for $\Pi_{\pm}$ are derived 
from those of $\Psi$ and $\bar \Psi$ via (\ref{pi_superfield}).

A formal solution of (\ref{super_flow_pi_field}) is given by 
\begin{eqnarray}
 \Pi_{\pm}(t,x)= \int d^4y \left\{ K_{t}^{\pm}(x-y)\Pi_{\pm}(0,y)
            + \int_0^t ds K_{t-s}^{\pm}(x-y) R_{\pm}(s,y)\right\},
\label{formal_solution_pi_field}
\end{eqnarray}
where $\theta,\bar\theta$ are abbreviated 
for $\Pi_{\pm}(t,x,\theta,\bar\theta)$ and $R_\pm(t,x,\theta,\bar\theta)$.
Here $K_t^{\pm}(x)$ is a heat kernel defined by 
\begin{eqnarray}
\begin{split}
 K_t^{\pm}(x)= \int \frac{d^4p}{(2\pi)^4} e^{ipx}e^{-t(p^2\mp im\sqrt{p^2})}.
\label{new_heat_kernel}
\end{split}
\end{eqnarray}
Note that (\ref{new_heat_kernel}) coincides with the normal one for $m=0$, 
and it still works as a damping factor  for $m\neq 0$.  
We can actually show that (\ref{formal_solution_pi_field}) satisfies (\ref{super_flow_pi_field})
because $K_t^{\pm}(x)$ provides a solution to the free part of  (\ref{super_flow_pi_field}).

We can also give the the formal of (\ref{super_flow_superfield}) as
\begin{eqnarray}
\begin{split}
&& \Psi_t(p) =  C_{t}(p)\Phi(p) - S_{t}(p)\frac{\bar D\bar D}{4\sqrt{p^2}}\bar\Phi(p)
\hspace{6cm}
   \\
&&  -\int_0^t ds \left( g^*C_{t-s}(p)\frac{\bar D\bar D}{4} (\bar\Psi_s\star\bar\Psi_s) (p)
                        +gS_{t-s}(p)\sqrt{p^2} (\Psi_s \star \Psi_s)(p)\right),\\
&& \bar \Psi_t(p)  = C_{t}(p)\bar\Phi(p) 
- S_{t}(p)\frac{DD}{4\sqrt{p^2}}\Phi(p)
\hspace{6cm}   
\\
&& -\int_0^t ds \left( g C_{t-s}(p)\frac{DD}{4} (\Psi_s\star\Psi_s)(p)
                        +g^* S_{t-s}(p)\sqrt{p^2} (\bar\Psi_s\star\bar\Psi_s)(p)\right),
\end{split}
\label{formal_solution_psi_field}
\end{eqnarray}
where we again abbreviate $\theta,\bar\theta$ of  
$\Psi_t(p,\theta,\bar\theta)$ and $\Phi(p,\theta,\bar\theta)$.
Here $D$ and $\bar D$ are the momentum representation of \eref{DD}, and  
 $C_t(p)$ and $S_t(p)$ are defined by
\begin{eqnarray}
C_t(p)&\equiv&e^{-tp^2}\cos(tm\sqrt{p^2}),\\
S_t(p)&\equiv&e^{-tp^2}\sin(tm\sqrt{p^2}),
\end{eqnarray}
which come from (\ref{new_heat_kernel}) in the momentum space as $K_t^{\pm} (p)=C_t (p)\pm iS_t (p)$. 
The star symbol means the convolution integral in the momentum space,
\begin{eqnarray}
(A\star B) (p) \equiv \int \frac{d^4q}{(2\pi)^4}A(q)B(p-q),
\end{eqnarray}
for any functions $A$ and $B$.
Note that $(A\star B)(p)=(B\star A)(p)$.

We finally find the formal solutions for the component fields 
inserting (\ref{Phi_expansion}) and (\ref{component_expression}) into (\ref{formal_solution_psi_field}),
\begin{eqnarray}
\phi_t(p)
       \aeq C_t(p)A(p)+\frac{i}{\sqrt{p^2}}S_t(p)F^{*}(p)
       -g\sqrt{p^2}\int^{t}_{0}dsS_{t-s}(p) (\phi_s\star\phi_s)(p)
\n\\
       &&+ g^* \int^{t}_{0}ds C_{t-s}(p) \left\{2i (\bar\phi_s \star \bar{G}_s)(p) - (\bar{\chi}_s \star \bar{\chi}_s)(p)\right\}, 
\\
 \bar\phi_t(p)
  \aeq C_t(p)A^*(p)+\frac{i}{\sqrt{p^2}}S_t(p)F(p) 
  -g^*\sqrt{p^2}\int^{t}_{0}ds S_{t-s}(p) (\bar{\phi}_s\star\bar\phi_s) (p)
\nonumber\\
&&
 +g\int^{t}_{0}ds C_{t-s}(p)\left\{2i (\phi_s\star G_s)(p) - (\chi_s\star\chi_s)(p) \right\},
\\   
\chi_{t}(p)
    \aeq C_t(p)\psi(p)-\frac{\sigma_{\mu} p_\mu}{\sqrt{p^2}}S_t(p)\bar{\psi}(p)
-2g^* \, \sigma_{\mu} p_\mu \int^{t}_{0}ds C_{t-s}(p) (\bar\phi_s\star\bar{\chi}_s) (p)\nonumber\\
&&-2g\sqrt{p^2}\int^{t}_{0}ds S_{t-s}(p)(\phi_s\star\chi_{s})(p),\\
\bar{\chi}_t(p)
\aeq C_t(p)\bar{\psi}(p) - \frac{\bar{\sigma}_{\mu} p_\mu}{\sqrt{p^2}}S_{t}(p)\psi(p)
-2g\,  \bar{\sigma}_\mu p_\mu \int^{t}_{0}ds C_{t-s}(p) (\phi_s\star\chi_{s})(p)\n\\
 &&-2g^*\sqrt{p^2}\int^{t}_{0}ds S_{t-s}(p) (\bar\phi_s\star\bar{\chi}_s)(p),\\
 G_t(p)
\aeq C_t(p)F(p)+i\sqrt{p^2}S_t(p)A^*(p) 
+ig^* p^2\int^{t}_{0}ds C_{t-s}(p) (\bar\phi_s\star\bar\phi_s)(p)\nonumber\\&&
-g\sqrt{p^2}\int^{t}_{0}ds S_{t-s}(p) \{2(\phi_s\star G_s)(p)+i (\chi_s\star\chi_s)(p)\},\\
\bar G_t(p)
\aeq C_t(p)F^{*}(p)+i\sqrt{p^2}S_t(p)A(p) 
+igp^2 \int^{t}_{0}ds C_{t-s}(p) (\phi_s\star\phi_s) (p)
\n\\
&& 
-g^*\sqrt{p^2}\int^{t}_{0}ds S_{t-s}(p)\{2 (\bar\phi_s\star\bar G_s)(p) + i (\bar{\chi}_s\star\bar{\chi}_s)(p)\}. 
\label{formal_solution_component_field}
\end{eqnarray}
Note that the terms with $1/{\sqrt{p^2}}$ are well-defined because 
they appear with $S_t(p)$
and $1/{\sqrt{p^2}}S_t(p)|_{p=0}=0$.

One of the interesting points is that the solutions 
have a damping oscillation with the flow time for nonzero mass, $C_t$ and $S_t$.
This behavior is different from the solution of the Yang-Mills flow whose damping factor is $e^{-t p^2}$.
In the case of $m=0$,  we have much simpler solutions because $C_t (p)=e^{-t p^2}$ and $S_t(p)=0$.

%% file: sec5.tex
\section{Summary} 
\label{sec:summary}

We have constructed a supersymmetric gradient flow equation 
in the four-dimensional Wess-Zumino model. 
The Wess-Zumino flow equation is given in two ways.
One is based on the off-shell component fields in which the flow for the scalar field is given by 
the gradient of the action. 
The flow equations for the other fields are derived from it by repeating the SUSY transformation.
The other way is based on the superfield formalism. 
The gradient flow for the chiral superfield 
is determined from the gradient of the action with respect to the superfield 
with keeping the supersymmetric chiral condition.
We found that the resultant equations are the same.

The obtained flow is supersymmetric in a sense that  the flow time derivative 
and the SUSY transformation commute with each other for nonzero flow time. 
On the other hand, the commutator does not vanish for the on-shell flow. 
The flowed components fields $G$ and $\bar{G}$ are not auxiliary but dynamical fields 
because the derivative terms are provided by their flows. 
We have obtained the formal solution of the Wess-Zumino flow equation 
and find that it behaves as a damping oscillation with respect to the flow time 
for nonzero mass,  
which is different from the Yang-Mills flow.

Since we have constructed the SUSY flow for the Wess-Zumino model, 
we achieved the first step 
toward the further understanding of the mechanism that leads to the UV finiteness 
of SUSY gradient flows. 
It is interesting whether the Wess-Zumino flow 
shows the UV finiteness at one loop order or not. In order to show that, 
further studies are now in progress.

%% file: app.tex
\section{Convention}
\label{sec:notation}

The Lorentz index $\mu$ runs $\mu=0,1,2,3$. 
All boundary fields are defined on ${\mathbb R}^4$.
The fermions $\psi_\alpha$ and $\bar\psi^{\dot\beta}$ transform
as spinors of $SO(4) \simeq SU(2)_L \times SU(2)_R$. 
The spinor indices $\alpha,\beta$ take the values $1,2$. 
We basically follow  \cite{Wess:1992cp} as the convention of spinors, but we 
perform the Wick rotation $t \rightarrow -it$ from \cite{Wess:1992cp}.
Then the auxiliary field $F$ is also replaced as $F \rightarrow iF$. 
Useful identities of a Euclidean (Wick rotated) version of  \cite{Wess:1992cp} 
are summarized in \cite{Kadoh:2018qwg}.

The antisymmetric tensors $\epsilon_{\alpha\beta}, \epsilon^{\alpha\beta},
\epsilon_{\dot\alpha\dot\beta},
\epsilon^{\dot\alpha\dot\beta}$ are defined as
$\epsilon_{21}=\epsilon^{12}=\epsilon_{\dot 2\dot 1}=\epsilon^{\dot 1\dot 2}=1$.
Spinors with upper and lower indices are defined as 
\begin{eqnarray}
\psi^\alpha=\epsilon^{\alpha\beta} \psi_\beta, \quad 
\bar\psi_{\dot\alpha} = \epsilon_{\dot\alpha\dot\beta} \bar\psi^{\dot\beta}.
\end{eqnarray}
Then Lorentz scalars made of two spinors are given by 
\begin{eqnarray}
\psi\chi \equiv \psi^\alpha\chi_\alpha, \quad 
\bar\psi\bar\chi \equiv \bar\psi_{\dot\alpha}\bar\chi^{\dot\alpha}.
\label{lorentz_scalar}
\end{eqnarray}
Note that $\bar\psi_{\dot\alpha}$ is not a complex conjugate of $\psi_{\alpha}$ in Euclidean space.

The four-dimensional 
sigma matrices $(\sigma_{\mu})_{\alpha\dot{\beta}}$ and $(\bar\sigma_{\mu})^{\dot\alpha \beta}$ 
are defined by  
\begin{eqnarray}
\sigma_0 = \bar\sigma_0 =-i \mathbf{1},  \quad \sigma_i=-\bar \sigma_i=\sigma^i,
\end{eqnarray}
where $\sigma^{i}$ for $i=1,2,3$ are the standard Pauli matrices.  
We often abbreviate the spinor index such as (\ref{lorentz_scalar}) throughout this paper. 
For instance, $\psi \sigma_{\mu} \bar{\psi}$ in the action (\ref{action_component})
means $\psi^\alpha (\sigma_{\mu})_{\alpha\dot\beta} \bar{\psi}^{\dot\beta}$.
The index structure of $\sigma_{\mu}$ and $\bar\sigma_{\mu}$ can be specified as  
$(\sigma_{\mu})_{\alpha\dot{\beta}}$ and $(\bar\sigma_{\mu})^{\dot\alpha \beta}$. 
They are related to each other as
\begin{eqnarray}
(\bar{\sigma}_{\mu})^{\dot{\alpha}\alpha} = \epsilon^{\dot\alpha\dot\beta}\epsilon^{\alpha\beta}
 (\sigma_{\mu})_{\beta\dot{\beta}}.
\end{eqnarray}
See \cite{Kadoh:2018qwg} for the other useful formulas after the Wick rotation.